\documentclass[aps,preprint]{revtex4-2}
\usepackage[english]{babel}
\usepackage[utf8]{inputenc}
\usepackage{physics}
\usepackage{caption}
\usepackage{amsmath}
\usepackage{amsthm}
\usepackage{graphicx}
\usepackage{subfigure}
\usepackage{threeparttable}
\usepackage{multirow}
\usepackage{dcolumn}
\usepackage{bm}
\usepackage{natbib}
\usepackage[colorlinks=true, allcolors=blue]{hyperref}
\usepackage{amssymb}
\everymath{\displaystyle}
\begin{document}

	\title{A Pure Quantum Approximate Optimization Algorithm \\ Based on CNR Operation}
	\author{Da You Lv}
	\author{An Min Wang}
	\altaffiliation{Corresponding author}
	\email{anmwang@ustc.edu.cn}
	\affiliation{Department of Modern Physics, University of Science and Technology of China, China}
\begin{abstract}
By introducing the ``comparison and replacement" (CNR) operation, we propose a general-purpose pure quantum approximate optimization algorithm and derive its core optimization mechanism quantitatively. The algorithm is constructed to a $p$-level divide-and-conquer structure based on the CNR operations. The quality of approximate optimization improves with the increase of $p$. For sufficiently general optimization problems, the algorithm can work and produce the near-optimal solutions as expected with considerably high probability. Moreover, we demonstrate that the algorithm is scalable to be applied to large size problems. Our algorithm is applied to two optimization problems with significantly different degeneracy, the Gaussian weighted 2-edge graph and MAX-2-XOR, and then we show the algorithm performance in detail when the required qubits number of the two optimization problems is 10.

\end{abstract}
\maketitle
\section{Introduction}\label{c1}
	Quantum computation can accelerate the approximate optimization and then promotes the development of related algorithms, which overcome the difficulty of exponential inefficiency hopefully. The well-known beginning is that Edward Farhi et. al developed the quantum adiabatic algorithm (QAA)\cite{farhi2000quantum} and proposed the quantum approximate optimization algorithm (QAOA)\cite{farhi2014quantum} which relies on the parameters produced by classical method. His team investigated applications of $p$-level QAOA in different combinatorial optimization problems such as typical instances\cite{Wang_2018,farhi2015quantum, lin2016performance,brandao2018fixed}, Sherrington-Kirkpatrick model\cite{Farhi_2022} and the ensemble of $k$-edge graphs\cite{Basso_2022}, etc. The results present some outstanding properties such as concentration\cite{Farhi_2022,Basso_2022}. And the performance of QAOA shows the quantum supremacy\cite{farhi2019quantum,Hadfield_2019,herrman2021multiangle} compared with classical algorithm. Moreover, QAOA has potential applications in a wide range of areas, including transportation science\cite{Azad_2023}, economy\cite{Amaro_2022}, product synthesis in biochemistry\cite{boulebnane2022peptide} and specific physics systems\cite{Ozaeta_2022,alaoui2020optimization,Chandarana_2022} etc. All these achievements motivate us to construct a general-purposed pure quantum approximate algorithm for combinatorial optimization problems in this paper.
	
	A combinatorial optimization problems can be quantified with cost function $C(z)$ and its arguments are defined as bits of the $n$-bit string $z=\left(z_1z_2\cdots z_n\right) \in\{0,1\}^n$\cite{Farhi_2022}. In this frame, the approximate optimization asks for a string $z^*$ such that $C(z^*)$ is as close to the absolute minimum $C_{min}$ as possible\cite{farhi2014quantum}.
	
	In this paper, we introduce the CNR operation to act on the tensor product of two states initially stored in two quantum registers, the target register $T_{qreg}$ and the support register $S_{qreg}$. Each register works in a $2^n$ dimensional Hilbert space spanned by $n$-qubit computational basis vectors $\{\ket{z}\}$, which is bit-wisely corresponding to $n$-bit strings $\{z\}$. And the cost function can be corresponded to a Hermitian operator $\mathcal{C}$ which is diagonal in the computational basis vectors, defined as
	\begin{equation}\label{eqn-3}
		\mathcal{C}\ket{z} = C(z)\ket{z}.
	\end{equation}
	The CNR operation is designed to indicate which register stores the string more optimum between $T_{qreg}$ and $S_{qreg}$, and if the string in $S_{qreg}$ is more optimum, the string in $T_{qreg}$ will be overwritten by the one in $S_{qreg}$ for each component of the superposition state. The procedure is accomplished with the aid of $t$ ancillary qubits. In the end, it produces a final state more optimum located in $T_{qreg}$ than the states initially stored in $T_{qreg}$ and $S_{qreg}$. 
	
	\begin{figure}[h]
		\centering
		\includegraphics[width=10cm]{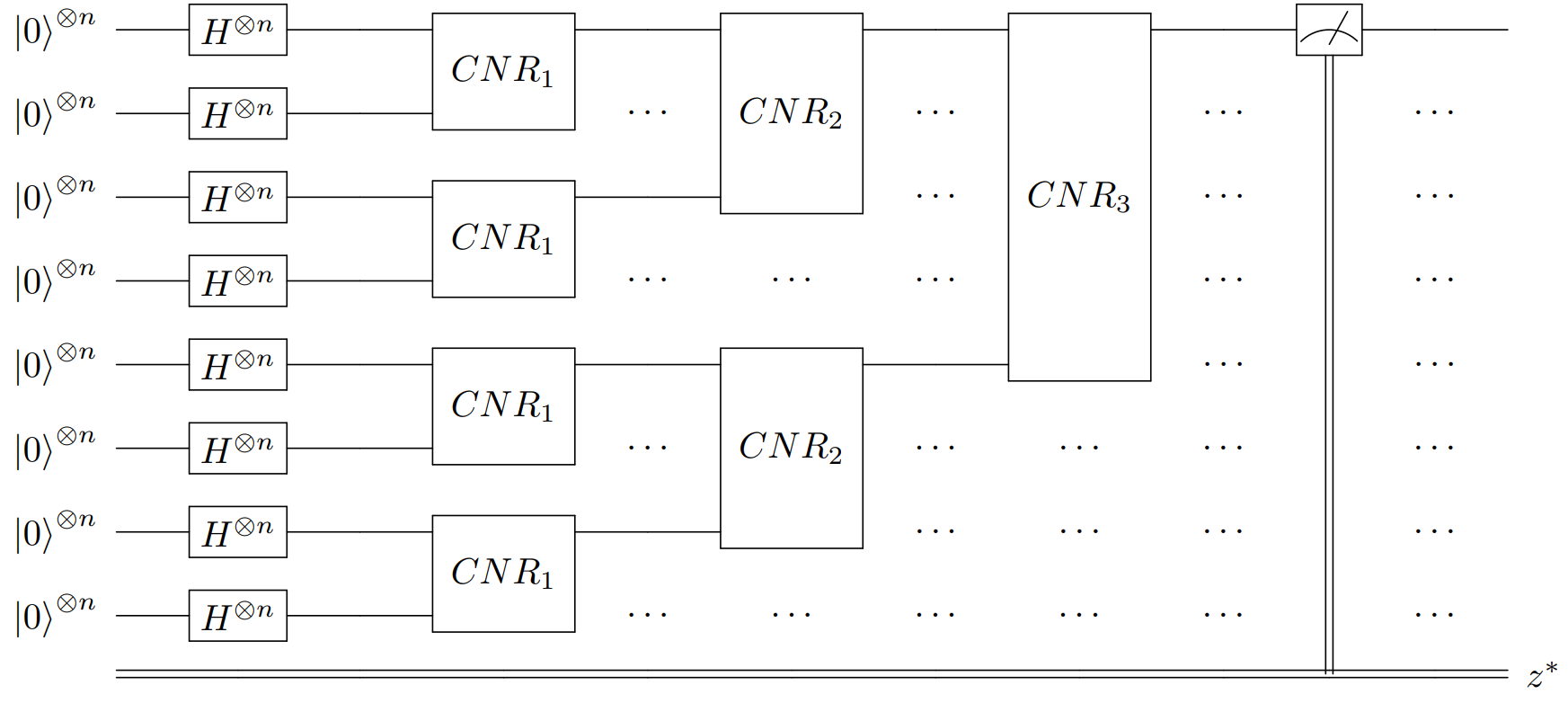}
		\captionsetup{justification=raggedright}
		\caption{\label{fig:divide_conquer}\sffamily {The divide-and-conquer structure of the 3-level algorithm. We omit the ancillary qubits of CNR and each quantum wire delivers $n$-qubit state for brevity.}}
	\end{figure}	
	
	Based on the CNR operation, we construct a quantum approximate optimization algorithm. The sketch of our algorithm when $p=3$ is shown in Figure \ref{fig:divide_conquer}. It presents the typical multi-level divide-and-conquer structure. There are $2^p$ registers in $\ket{0}^{\otimes n}$ as the input of the $p$-level algorithm. Then they will be transformed into $n$-qubit uniform superposition by Hadamard gates and enter the first level as $T_{qreg}$ and $S_{qreg}$ of CNR operations by pairs. The input of the $(m+1)$-th level are $2^{p-m}$ target registers from the $m$-th level. These registers are divided into new target-support register pairs as the input of CNR in the $(m+1)$-th level. Then $2^{p-m-1}$ CNR operations in the $(m+1)$-th level output totally $2^{p-m-1}$ target registers to the $(m+2)$-th level. By repetitions of the CNR operation, the quality of approximate optimization improves level by level. Finally, the $p$-level algorithm outputs the state in the last $T_{qreg}$ experiencing all $p$-level CNR operations. The last step is measuring the state to obtain a string $z^*$. According to the algorithm procedure, the $p$-level algorithm needs $O(2^p(n+t))$ qubits (including ancillary qubits). For sufficiently general optimization problems, the algorithm shown above can work and produce the near-optimal solutions as expected with considerably high probability.
	
	In this paper, we propose a pure quantum approximate optimization algorithm based on the CNR operations. By performing the CNR operation, we quantitatively derive the core optimization mechanism in sufficiently general problems. In order to discuss the algorithm performance and analysis the quality of approximation under the uncertainty of quantum computation, we introduce the relative error as a measurement of precision and clearly define the proper derived quantities to finish the analysis about algorithm performance. In particular, with the help of these conception, we demonstrate the scalability of our algorithm, which means that the algorithm can be applied to the problems with large size and produce the near-optimal solutions in principle. As an investigation of the algorithm performance, we apply the algorithm to two important instances with different degeneracy when $n=10$ and illustrate the results including the cumulative probability over the near-optimal solutions and the relative error to the optimal solution.

	The paper is organized as follows. In section \ref{c2}, we put forward the CNR operation by realizing two sub-operations, comparison and replacement in detail. Then we analysis the theoretical algorithm performance in section \ref{c3}, where we derive the problem-independent relation to calculate the probability distribution of measurement, analyze the quantum algorithm performance and demonstrate the scalability of our algorithm. Subsequently, we apply the algorithm to the Gaussian weighted 2-edge graph and MAX-2-XOR when $n=10$ in section \ref{c4}, where we reorganize the obtained results and illustrate the algorithm performance properly. As a conclusion, we review the algorithm and emphasize the results of theoretical analysis and actual application, and then finish some further discussions in section \ref{c5}.
	
	\section{Comparison and Replacement}\label{c2}
		
		\subsection{Comparison operation with ancillary register}
		We start by introducing the first step of CNR, the comparison operation. It compares the two strings located in $T_{qreg}$ and $S_{qreg}$ respectively in sense of optimization and stores the answer information in the ancillary register. It can be implemented by the similar technique in Quantum Phase Estimation (QPE)\cite{Cleve_1998}. The input of comparison is the $2n$-qubit state composed by tensor product of $T_{qreg}$ and $S_{qreg}$. We introduce $t$ qubits in the uniform superposition state as the ancillary register. The operator of the eigenvalue estimation is $\mathcal{A} = I\otimes \mathcal{C} - \mathcal{C}\otimes I$, where $\mathcal{C}$ is the Hermitian operator defined by (\ref{eqn-3}). The corresponding unitary operator is just
		\begin{equation}\label{eqn-9}
			\exp\left(\mathrm{i}\frac{\mathcal{A}}{M}\right) = \exp\left(\mathrm{i}\frac{\mathcal{C}_s}{M}\right)\otimes \exp\left(-\mathrm{i}\frac{\mathcal{C}_t}{M}\right),
		\end{equation}
		where $\mathcal{C}_s$ or $\mathcal{C}_t$ only acts on $S_{qreg}$ or $T_{qreg}$ respectively. And the scale factor $M$ is introduced to scale the spectrum of $\mathcal{A}$ in the range $[-\pi,\pi)$ to avoid the multi-value correspondence from the periodicity of exponent on the imaginary axis. Thus the strict lower bound that $M$ needs to satisfy is
		\begin{equation}\label{eqn-10}
			M \geq \frac{C_{max} - C_{min}}{2\pi}
		\end{equation}
		The exact bound of $M$ for a general problem is unknown, unless the problem is solved. But a suitable $M$ can be estimated, since the number of edges in a graph has an upper bound and coefficients can be estimated according to some feature quantities, such as the regular of vertex in MAX-k-XOR or $(\mu,\sigma)$ for Gaussian weighted graphs. 
		
		When the comparison acts on a tensor product component $\ket{z_t}\ket{z_s}$, where $\ket{z_t}$ and $\ket{z_s}$ are located in $T_{qreg}$ and $S_{qreg}$ respectively, the operation can be described as follows. It first assigns $\displaystyle{\frac{C(z_s)-C(z_t)}{M}}$ as phase on the corresponding tensor product, by controlled-$\exp\left(\mathrm{i}2^j\frac{\mathcal{A}}{M}\right)$ according to ancillary qubits in uniform superposition. Then the inverse Quantum Fourier Transformation on total $2n+t$ qubits extracts the information in phases into the ancillary register. The procedure of comparison in explicit expression is
		\begin{equation}\label{eqn-8}
			\ket{z_t}\ket{z_s}\sum_{x\in\{0,1\}^t}\frac{1}{\sqrt{2^t}}\ket{x}\rightarrow \frac{1}{\sqrt{2^t}}\sum_{x\in\{0,1\}^t} e^{\mathrm{i}2\pi D(x) \Delta }\ket{z_t}\ket{z_s}\ket{x}\rightarrow \ket{z_t}\ket{z_s}\ket{\tilde{\Delta}},
		\end{equation}
		where $\{\ket{x}=\ket{x_1x_2\cdots x_t}\}$ are computational basis vectors working for the ancillary register. Here, we use $D(x)$ to denote the decimal value of string $x$ and $\Delta$ to refer to $\displaystyle{\frac{C(z_s)-C(z_t)}{2\pi M}}$ in the range $\left[-\frac{1}{2},\frac{1}{2}\right)$.

\begin{figure}[h]
			\centering
			\subfigure{(a)}{
				\includegraphics[width=7cm]{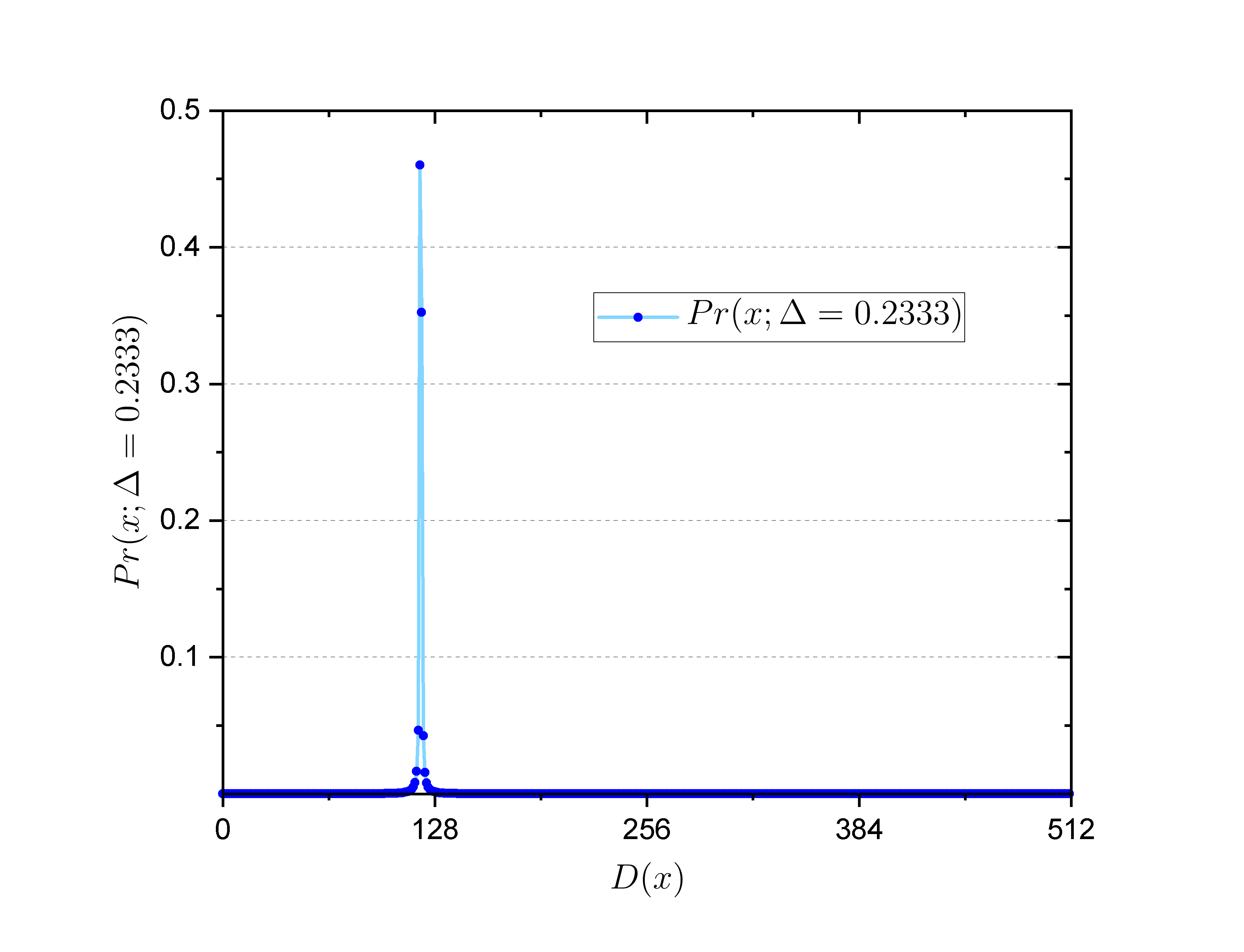}
				\label{fig: prposi}
			}
			\subfigure{(b)}{
				\includegraphics[width=7cm]{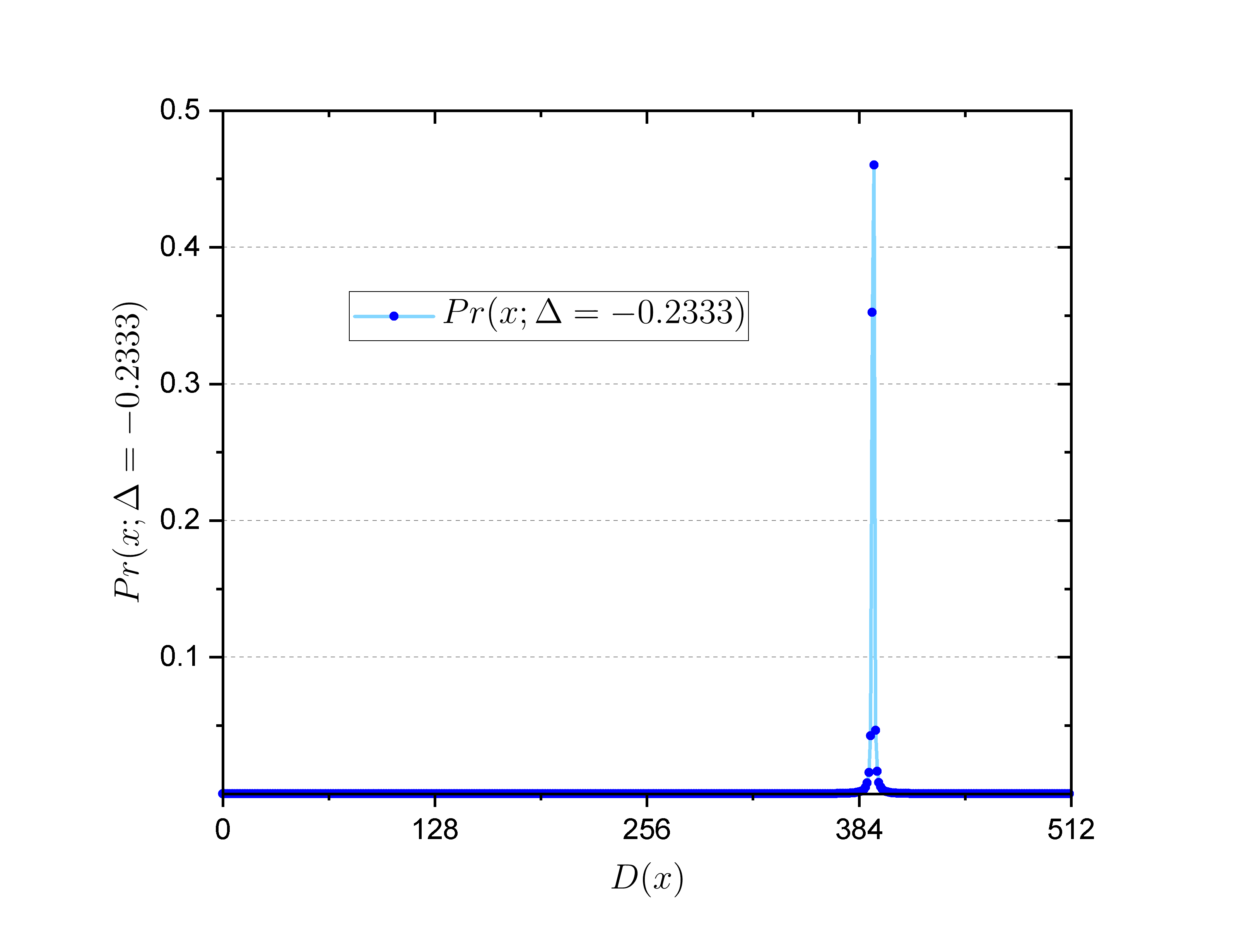}
				\label{fig: prnega}
			}
			\captionsetup{justification=raggedright}
			\caption{\label{fig:true distribution}\sffamily{(a) The probability distribution of measuring $\ket{\tilde{\Delta}}$, where $\Delta = 0.2333$ and $t = 9$. (b) The probability distribution of measuring $\ket{\tilde{\Delta}}$, where $\Delta = -0.2333$ and $t = 9$. }}
		\end{figure}

		By the expression (\ref{eqn-8}), the final state of comparison consists of the tensor product $\ket{z_t}\ket{z_s}$ ,as well as $\ket{\tilde{\Delta}}$ which stores the information of $\Delta$. Expanding $\ket{\tilde{\Delta}}$ with $t$-qubit computational basis vectors, the amplitude of $\ket{x}$ is\cite{Cleve_1998}
		\begin{equation}\label{eqn-14}
			\phi(x;\Delta) = \frac{1}{2^t} \frac{1-\exp\left(\mathrm{i}2\pi\left(2^t\Delta-D(x)\right)\right)}{1-\exp\left(\mathrm{i}2\pi\left(\Delta-2^{-t}D(x)\right)\right)}.
		\end{equation}
		And the corresponding probability is
		\begin{equation}\label{eqn-7}
			Pr(x;\Delta) = \frac{1}{2^{2t}} \frac{1-\cos(2\pi(2^t\Delta-D(x)))}{1-\cos\left(2\pi\left(\Delta-2^{-t}D(x)\right)\right)},
		\end{equation}
		which peaks at $x$ such that $D(x)=2^t\Delta$ when $\Delta\geq0$ or $D(x)=2^t(\Delta+1)$ when $\Delta <0$. Figure \ref{fig:true distribution}(a) and (b) illustrate the probability distribution for the two cases separately. We especially care about whether $\Delta$ is positive or negative, which determines if the replacement operation introduced later is triggered. It can be seen that in the ancillary register, components corresponding to $D(x)<2^{t-1}-1$ has the first ancillary qubit $\ket{0}$, which stand for $C(z_s)-C(z_t)\geq 0$, that is, $C(z_t)$ is closer to the $C_{min}$. And conversely those components corresponding to $D(x)\geq2^{t-1}$ has the first ancillary qubit $\ket{1}$, which gives a opposite result $C(z_s)-C(z_t)<0$, that is $C(z_s)$ is closer to the $C_{min}$
		
		In application, presetting an accuracy can effectively reduce the cost of resource. In our algorithm, when the comparison operation gives an incorrect answer for the sign of $\Delta$, we count the case as a failure only when $|\Delta|$ is larger than accuracy $b$. For MAX-k-XOR or MAX-k-SAT, the absolute value of gap between different cost function values is at least 2, thus we can choose $\frac{2}{M}$ as accuracy for the best performance. However when the coefficients are in continuous distribution, pursuing performance is costly. There exists a trade-off between performance and efficiency. When we choose a proper accuracy, $t$ can be determined by $t\geq \text{log}_{2}b$ to ensure that the comparison works as expected.
	
	\subsection{Replacement controlled by the first ancillary qubit}
	In the implementation of comparison, one point worth emphasizing is that the answer of comparing the two string can be delivered by the first qubit in our ancillary register. Thus, more explicitly, as the second step of CNR, the replacement operation is asked that for the first ancillary qubit at $\ket{1}$, it replaces the computational basis vector in $T_{qreg}$ with the one in $S_{qreg}$, and for the first ancillary qubit at $\ket{0}$, it does nothing. 
	
\begin{figure}[h]
		\centering
		\includegraphics[width=8.5cm]{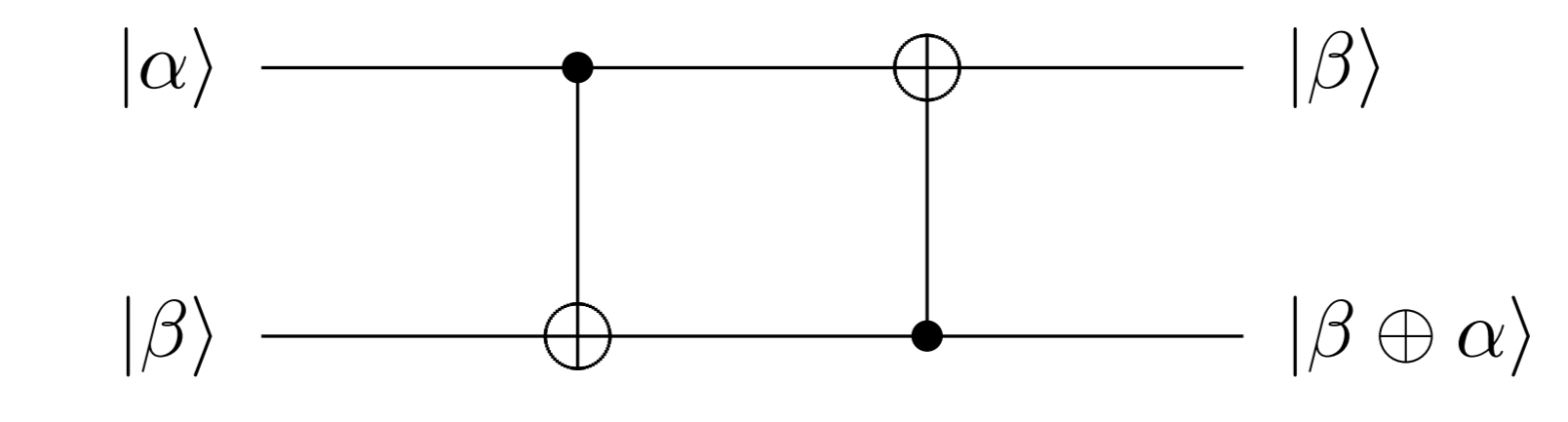}
		\caption{\label{fig:overwriting}\sffamily{The quantum circuit of single-qubit overwriting operation.}}
	\end{figure}

	The replacement operation can be implemented directly by controlled-overwriting operation introduced by us. The quantum circuit of single-qubit overwriting is shown in Figure \ref{fig:overwriting}.
	We use the first ancillary qubit from comparison as the control bit, and construct the bit-wise controlled-overwriting between two $n$-qubit states, by $n$ simultaneous single-qubit overwriting operations between qubits at the same position in $T_{qreg}$ and $S_{qreg}$ correspondingly. Such a operation meets our requirement above to work as the second sub-operation after the comparison.
	
	So far, we realize the CNR operation by performing the comparison and the replacement in turn. Since both comparison and replacement are unitary, by quantum parallelism, CNR still works when the input is a $2n$-qubit state stored jointly by $T_{qreg}$ and $S_{qreg}$. For a general input, each tensor product component will experience the procedure we introduced before. The CNR operation can transform the initial state and output a final state as expected.

	\section{Theoretical algorithm performance}\label{c3}
	As a quantum approximate optimization algorithm, our algorithm obtains the complexity improvement mainly from the quantum computation. However, we must investigate the obtaining probability of the required approximate solution, the quality of approximate optimization and whether the algorithm can be applied to optimization problems with large size $n$, etc.
	 
	We first introduce and explain the labels and notes used in the following context. The Hermitian operator  $\mathcal{C}$ can be understood as a Hamiltonian. In general, the correspondence between strings and cost function values is not simply one-to-one. Thus in terms of the Hamiltonian, one energy level can include multiple computational basis vectors as energy eigenstates, in other words the energy level is degenerate. According to equation (\ref{eqn-3}), we can relabel computational basis vectors in the complete-knowing sight over the problem. To keep the generality, we use $\ket{E_{i;j}}$ to denote the $j$-th eigenstate in the subspace of energy $E_i$ which is corresponding to the $i$-th energy level. It must be declared here that the complete knowledge is not necessary for CNR to optimize general combinatorial optimization problems. The only requirement is a black box that can perform the controlled-$\mathrm{exp}\left( \mathrm{i}j\frac{\mathcal{C}}{M}\right)$ operation for integer $j$. Therefore, this relabeling on computational basis vectors will not change the objective result. We assume that there are $G$ energy levels and the $i$-th level includes $g_i$ eigenstates.
	
	Now, let us to see how the CNR operation produces the final state from the general tensor product of states in $T_{qreg}$ and $S_{qreg}$ initially. Without loss of generality, we consider the initial states in $T_{qreg}$ and $S_{qreg}$ as
	\begin{equation}\label{eqn-23}
		\ket{T} = \sum_{i}^{G}\sum_{j=1}^{g_i}c_{i;j}\ket{E_{i;j}},
	\end{equation}
	\begin{equation}\label{eqn-24}
		\ket{S} = \sum_{k=1}^{G}\sum_{l=1}^{g_k}  q_{k;l}\ket{E_{k;l}}.
	\end{equation}
	where $\{c_{i;j}\}$ and $\{q_{k;l}\}$ are coefficients that have been normalized. By our definition in section \ref{c2}, the action of CNR operation can be written explicitly as follows
	\begin{equation}\label{eqn-103}
		\begin{split}
			\sum_{i,k=1}^{G}\sum_{j=1}^{g_i}\sum_{l=1}^{g_k} c_{i;j}q_{k;l}\ket{E_{i;j}}\ket{E_{k;l}}\sum_{x\in\{0,1\}^n}\frac{1}{\sqrt{2^t}}\ket{x}\ \ \ \ \ \ \ \ \ \ \ \ \ \ \ \ \ \ \ \ \ \ \ \ \ \ \ \ \  \\
			\rightarrow  \sum_{i,k=1}^{G}\sum_{j=1}^{g_i}\sum_{l=1}^{g_k} c_{i;j}q_{k;l}\ket{E_{i;j}}\ket{E_{k;l}}\sum_{x\in\{0,1\}^n;x_1=0} \phi(x;\Delta_{i,k}) \ket{x}  \\ 	+\sum_{i,k=1}^{G}\sum_{j=1}^{g_i}\sum_{l=1}^{g_k} c_{i;j}q_{k;l}\ket{E_{k;l}}\ket{E_{k;l}\bar{\oplus} E_{i;j}}\sum_{x\in\{0,1\}^n;x_1=1}\phi(x;\Delta_{i,k}) \ket{x},
		\end{split}
	\end{equation}
	where $\ket{E_{k;l}\bar{\oplus} E_{i;j}}$ is defined as the state produced by bit-wise XOR between $\ket{E_{i;j}}$ and $\ket{E_{k;l}}$. And we define $\displaystyle{\Delta_{i,k}=\frac{2^t(E_k-E_i)}{2\pi M}}$ for brevity. We assume $t$ has been chosen properly. The comparison in CNR gives a correct answer through the first ancillary qubit with probability close to 1. We omit those non-peaking terms in the summation, and normalize the expression again, thus (\ref{eqn-103}) becomes
	\begin{equation}\label{eqn-102}
		\begin{split}
	\sum_{i,k=1}^{G}\sum_{j=1}^{g_i}\sum_{l=1}^{g_k} c_{i;j}q_{k;l}\ket{E_{i;j}}\ket{E_{k;l}}\sum_{x\in\{0,1\}^n}\frac{1}{\sqrt{2^t}}\ket{x} \\ \rightarrow 
	\sum_{i\leq k}^{G}\sum_{j=1}^{g_i}\sum_{l=1}^{g_k}c_{i;j}q_{k;l}\ket{E_{i;j}}\ket{E_{k;l}}\ket{0\cdots} \\ +  \sum_{i>k}^{G}\sum_{j=1}^{g_i}\sum_{l=1}^{g_k}c_{i;j}q_{k;l}\ket{E_{k;l}}\ket{E_{k;l}\bar{\oplus} E_{i;j}}\ket{1\cdots},
		\end{split}
	\end{equation}
	where we highlight the first ancillary qubit $x_1$ and omit the rest $t-1$ qubits in ancillary register using the ellipse. Since the terms after the arrow are orthogonal to each other, the probability of the $a$-th energy level from $T_{qreg}$ after CNR operation is
	\begin{equation}\label{eqn-25}
		\begin{split}
			Pr^{CNR}(a) =& \sum_{k\geq a}\sum_{j=1}^{g_a}\sum_{l=1}^{g_k}|c_{a;j}q_{k;l}|^2 + \sum_{i> a}\sum_{j=1}^{g_i}\sum_{l=1}^{g_a}|c_{i;j}q_{a;l}|^2 \\
			=& \sum_{j=1}^{g_a}|c_{a;j}|^2 + \sum_{l=1}^{g_a}|q_{a;l}|^2 
			- \sum_{i\leq a}\sum_{j=1}^{g_i}\sum_{l=1}^{g_a} |c_{i;j}q_{a;l}|^2 \\
			-& \sum_{j=1}^{g_a}\sum_{k\leq a}\sum_{l=1}^{g_k} |c_{a;j}q_{k;l}|^2 
			+ \sum_{j=1}^{g_a}\sum_{l=1}^{g_a}|c_{a;j}q_{a;l}|^2 \\
			=& Pr(a)+{Pr}^{\prime}(a) - Pr(a)\sum_{i=1}^{a}{Pr}^{\prime}(i) - {Pr}^{\prime}(a)\sum_{i=1}^{a}Pr(i) + Pr(a){Pr}^{\prime}(a).
		\end{split}
	\end{equation}
	In a degenerate energy level, eigenstates give equivalently approximate solutions. Thus the probability of obtaining energy levels from the final state is more general and worth to be concerned. In the last step of (\ref{eqn-25}), we organize the result according to (\ref{eqn-23}) and (\ref{eqn-24}), and denote the probability of obtaining the $a$-th energy level as $Pr(a)$ for $\ket{T}$, and ${Pr}^{\prime}(a)$ for $\ket{S}$. And it shows the general relation between the final state in $T_{qreg}$ and initial states in $T_{qreg}$ and $S_{qreg}$ for measurement probability of energy levels.
	
	Then we apply the relation (\ref{eqn-25}) to our algorithm. According to the algorithm structure shown in Figure \ref{fig:divide_conquer}, each CNR operation in the algorithm receives $S_{qreg}$ identical to $T_{qreg}$. Thus we can simply remove the primes in (\ref{eqn-25}). For the CNR operation in the $(m+1)$-th level, both $T_{qreg}$ and $S_{qreg}$ are the output states from the $m$-th level of CNR operations. We denote the probability of the $a$-th energy level from the output state of the CNR operation in the $m$-th level in our algorithm as $P(a,m)$, and the summation of probability from the first energy level to $a$-th level as $S(a,m)$. For completeness of theory, we introduce $m=0$ to refer to the $n$-qubit uniform superposition state. Remove the primes in (\ref{eqn-25}) and plug $P(a,m)$ and $S(a,m)$ into it. We obtain the recursion relation between final state in $T_{qreg}$ of a CNR operation in the $m$-th level and $(m+1)$-th level
	\begin{equation}\label{eqn-108}
		P(a,m+1) = 2P(a,m) - 2P(a,m)S(a,m) + P(a,m)^2.
	\end{equation}
	By summing both sides over the first energy level to the $a$-th level above, we have a more convenient form
	\begin{equation}\label{eqn-109}
		S(a,m+1) = 1 - (1-S(a,m))^2 = 1-(1-S(a,0))^{2^{(m+1)}}.
	\end{equation}
	It is reliable for sufficiently general optimization problems, since our derivation above is in the general form. Combining (\ref{eqn-109}) and the input in the uniform superposition state, we can calculate the probability distribution on energy levels for our algorithm with $p\geq0$. It is (\ref{eqn-109}) that reveals the optimization mechanism of our algorithm.
	
	Because of the intrinsic uncertainty of quantum algorithm, it is necessary to consider all possible near-optimal solutions and their degree of approximation to the optimal solution. Also, we care about whether these solutions is obtainable according to the measurement probability. We quantize the quality of approximation by introducing the relative error that depends on a string $z$ for the minimum value problems, which is defined as
	\begin{equation}\label{eqn-4}
		\alpha_R(z) = \frac{C(z)-C_{min}}{C_{max}-C_{min}}.
	\end{equation}
	It should be pointed out that in order to avoid the difficulty of possible zeros in the denominator to describe sufficiently general optimization problems, we adopt the definition of relative error as a measurement of precision with respect to the ranges of cost function, which is denoted as by the subscript $R$. According to the definition, as $C(z)$ is closer to $C_{min}$, the $\alpha_R(z)$ is closer to 0, which means the string $z$ is more optimum. Since $\alpha_R(z)$ is constant within each energy level, we can also talk about the relative error that depends on the $a$-th energy level and simply use $\alpha_R(a)$ to denote it.

	We introduce the worst relative error bound $\epsilon_R$ to put the requirement on algorithm to find a string $z$ such that $\alpha_R(z)\leq\epsilon_R$. The worst relative error bound forces us to concern about those near-optimal solutions satisfying the requirement within the neighborhood of the ground energy level $U(1,\beta)$. It is the first $\beta+1$ energy levels, and $\beta$ as the width of neighborhood is an integer larger than 0. For a problem sized $n$, the cumulative probability over $U(1,\beta)$, denoted as $\mathcal{P}(U(1,\beta),p)$, can be evaluated directly by (\ref{eqn-109}). Without loss of generality, we assume the energy levels in $U(1,\beta)$ consists of $A_{\beta}$ computational basis vectors, where $A_{\beta}$ is completely determined by the condition of degeneracy. And it is straightforward that $A_{\beta}\geq\beta+1$, where the equal sign is established if and only if the first $\beta+1$ energy level is nondegenerate. The cumulative probability is evaluated according to the definition in general
	\begin{equation}\label{eqn-112}
		\mathcal{P}(U(1,\beta),p) = S(\beta+1,p) = 1-\left( 1-\frac{A_{\beta}}{2^n}\right) ^{2^p},
	\end{equation}
	Consider the function defined on the positive integer $N$ with a positive real number $c$
	\begin{equation}\label{eqn-113}
		F(N) = \left( 1-\frac{1}{N}\right)^{cN}.
	\end{equation}
	It increases monotonically with the growth of $N$, and converges to $\frac{1}{\mathrm{e}^c}$ when $N$ trends to infinity. Thus
	\begin{equation}\label{eqn-114}
		1-\left( 1-\frac{1}{N}\right)^{cN} \geq 1-\frac{1}{\mathrm{e}^c}.
	\end{equation}
	is established. Plug the right side of (\ref{eqn-112}) into (\ref{eqn-114}), and obtain a lower bound of cumulative probability over $U(1,\beta)$
	\begin{equation}\label{eqn-115}
		\mathcal{P}(U(1,\beta),p)=1-\left( 1-\frac{A_{\beta}}{2^n}\right) ^{2^p}= 1-\left( 1-\frac{A_{\beta}}{2^n}\right) ^{\frac{2^n}{A_{\beta}}\frac{A_{\beta}}{2^n}2^p} \geq 1-\mathrm{exp}\left( -2^p\frac{A_{\beta}}{2^n}\right). 
	\end{equation} 
	If we expect that
	\begin{equation}\label{eqn-116}
		\mathcal{P}(U(1,\beta),p) \geq \eta,
	\end{equation}
	where $\eta \in [0,1]$ is the probability of acceptance preset by us, we can solve the lower bound for $p$
	\begin{equation}\label{eqn-140}
		p \geq \log_2{\left(\frac{2^n}{A_{\beta}} \ln{\left( \frac{1}{1-\eta}\right)}\right)}
	\end{equation}
	
	The cumulative probability over $U(1,\beta)$ converges to 1 very quickly as $p$ increases according to (\ref{eqn-115}), thus the solutions asked by the preset bound $\epsilon_R$ is obtainable. Based on the discussion above, we can also evaluate the average relative error on $U(1,\beta)$, denoted as $\bar{\alpha}_R(U(1,\beta))$, to characterize the average performance in those near-optimal solutions, which is defined as
	\begin{equation}\label{eqn-141}
			\bar{\alpha}_R(U(1,\beta)) = \frac{1}{\mathcal{P}(U(1,\beta),p)}\sum_{i=1}^{\beta+1}P(i,p)\alpha_R(i).
	\end{equation}
	And using (\ref{eqn-109}) to evaluate $P(i,p)$, $\bar{\alpha}_R(U(1,\beta))$ can be calculated in principle.

	We now demonstrate the scalability of our algorithm.
	\begin{figure}[h]
		\centering\subfigure{(a)}{
			\includegraphics[width=7cm]{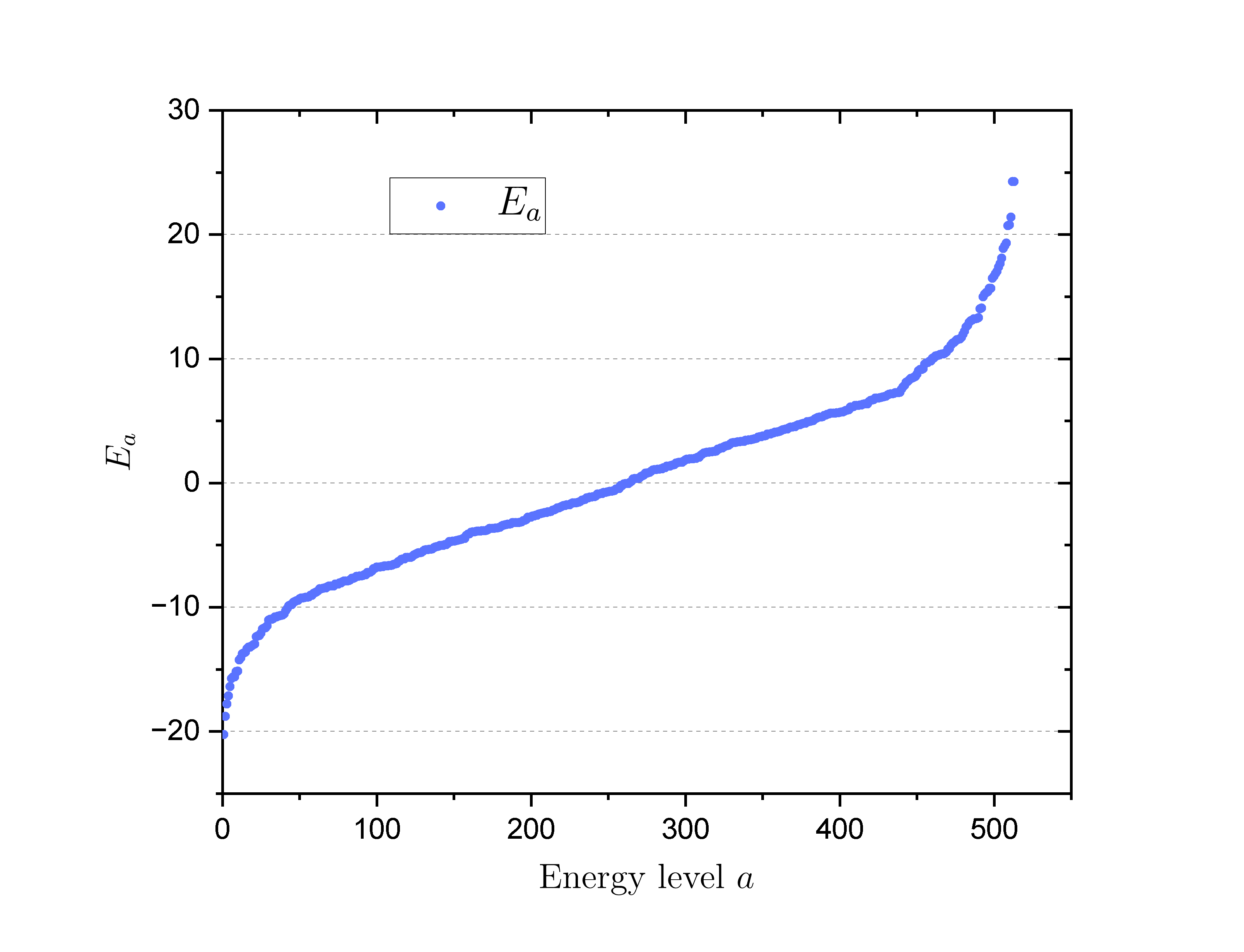}
		}
		\subfigure{(b)}{
			\includegraphics[width=7cm]{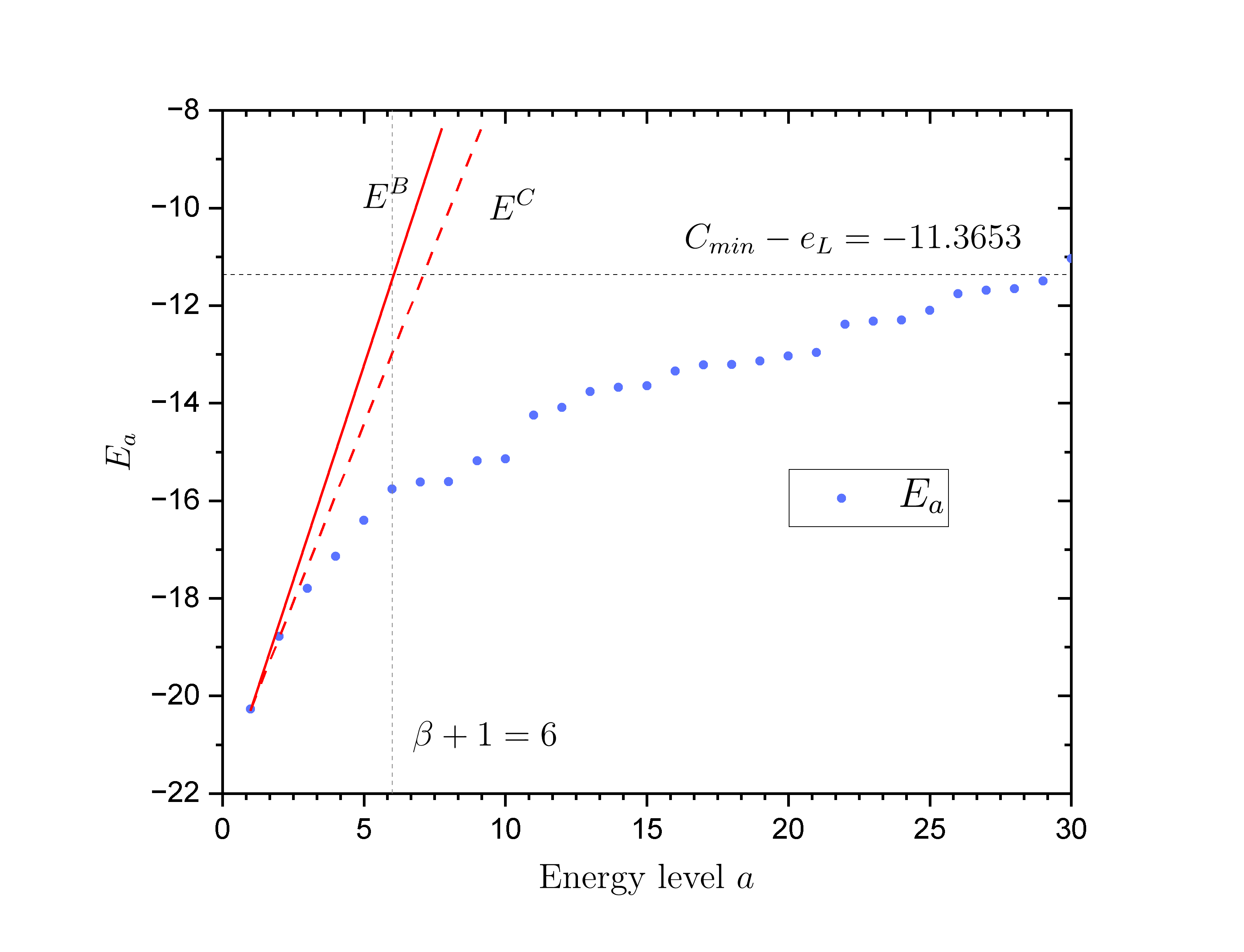}
		}
		\captionsetup{justification=raggedright}
		\caption{\label{fig:upperline_gauss}\sffamily{(a) The scatter graph of energy corresponding to energy levels in ascending order for the Gaussian weight 2-edge graph. (b) The locally enlarged graph of the first several energy levels for the energy curve $E_a$ of the Gaussian weighted 2-edge graph, the critical linear function $E^C(a)$ and a upper bound straight line $E^B(a)$ in condition that $\epsilon_R=0.2$.  } }
	\end{figure}
	Without loss of generality, we take the problem whose energy changing graph along with energy levels is shown in Figure \ref{fig:upperline_gauss}(a) as example. There exists a critical upper bound straight line of $E_a$ described by the linear function 
	\begin{equation}\label{eqn-402}
		E^C(a)=E_1 + \frac{E_{\tilde{a}}-E_1}{\tilde{a}-1}(a-1),
	\end{equation}
	where $\tilde{a}$ is the energy level such that the slope from $(1,E_1)$ to $(\tilde{a},E_{\tilde{a}})$ is largest. It is obvious that $E^C(a)\geq E_a$. The existence of $E^C(a)$ is sufficient general, since we put no restriction on the problem size, degeneracy or energy. In principle, there are infinite many upper bound straight lines of $E_a$ starting at the point $(1,E_1)$. $E^C(a)$ is the one with the lowest slope. And we denote the upper bound linear functions as $E^B(a)$. 

	We assume that a linear function of a upper bound straight line is
	\begin{equation}\label{eqn-405}
		E^B(a)=E_1 + \frac{\epsilon_R(C_{max}-C_{min})}{\beta}(a-1) \geq E^C(a).
	\end{equation}
	Replacing $E_a$ with the linear function $E^B(a)$, the worst relative error up to the $\beta+1$ energy level is just $\epsilon_R$. Since when $a\geq1$, $E^B(a)\geq E^C(a)\geq E_a$, using $E^B(a)$ gives a worse relative error than the true energy. Thus the existence of upper bound straight line can be written in form of (\ref{eqn-405}) is corresponding to the existence of a neighborhood $U(1,\beta)$ where the bound $\epsilon_R$ can be satisfied. At the same time, $E^B(a)\geq E^C(a)$ proposes a upper bound of $\beta$ that
	\begin{equation}\label{eqn-406}
		\beta \leq \left\lfloor\frac{\epsilon_R(C_{max}-C_{min})(\tilde{a}-1)}{E_{\tilde{a}}- E_1}\right\rfloor.
	\end{equation} 
	According to (\ref{eqn-406}), if we ask for a better approximation (lower $\epsilon_R$), we need to reduce the width of neighborhood $U(1,\beta)$. And we allow $\beta=0$ to correspond to a vertical line with infinite slope, which is the case that we only concern the ground energy for those extremely small $\epsilon_R$. In this condition, there must exist a $U(1,\beta)$ concerned by us such that the worst bound $\epsilon_R$ can be satisfied for sufficiently general optimization problems. Figure \ref{fig:upperline_gauss}(b) is the local enlarged graph showing the true energy, the critical upper bound straight line $E^C(a)$ and $E^B(a)$ corresponding to $\epsilon_R=0.2$.
	
	By (\ref{eqn-115}), we can evaluate the cumulative probability over $U(1,\beta)$
	\begin{equation}\label{eqn-407}
		\mathcal{P}(U(1,\beta),p)=1-\left(1-\frac{A_{\beta}}{2^n}\right)^{2^p} \geq 1-\mathrm{exp}\left( -2^p\frac{A_{\beta}}{2^n}\right). 
	\end{equation}
	For fixed $n$, no matter how large it is, we can always choose a $p$ to make the right side of the inequality sufficiently close to 1. For example, even considering an extreme case that $A_{\beta}=1$, the cumulative probability is higher than $1-\frac{1}{\mathrm{e}^8} \approx 0.9997$ when $p=n+3$. Though such a large $p$ is no significance in actual applications since the spatial complexity exponentially relies on $p$, we demonstrate that there always exists a $p$ such that the cumulative probability over the solutions satisfying the requirement close enough to 1 in principle. The existence of $U(1,\beta)$ for any $\epsilon_R$ and the corresponding $p$ to amplify the cumulative probability to sufficiently close to 1 fully demonstrates the scalability of our algorithm.

\section{Application in instances}\label{c4}
	\begin{figure}[h]
	\centering
	\subfigure{(a)}{
		\includegraphics[width=7cm]{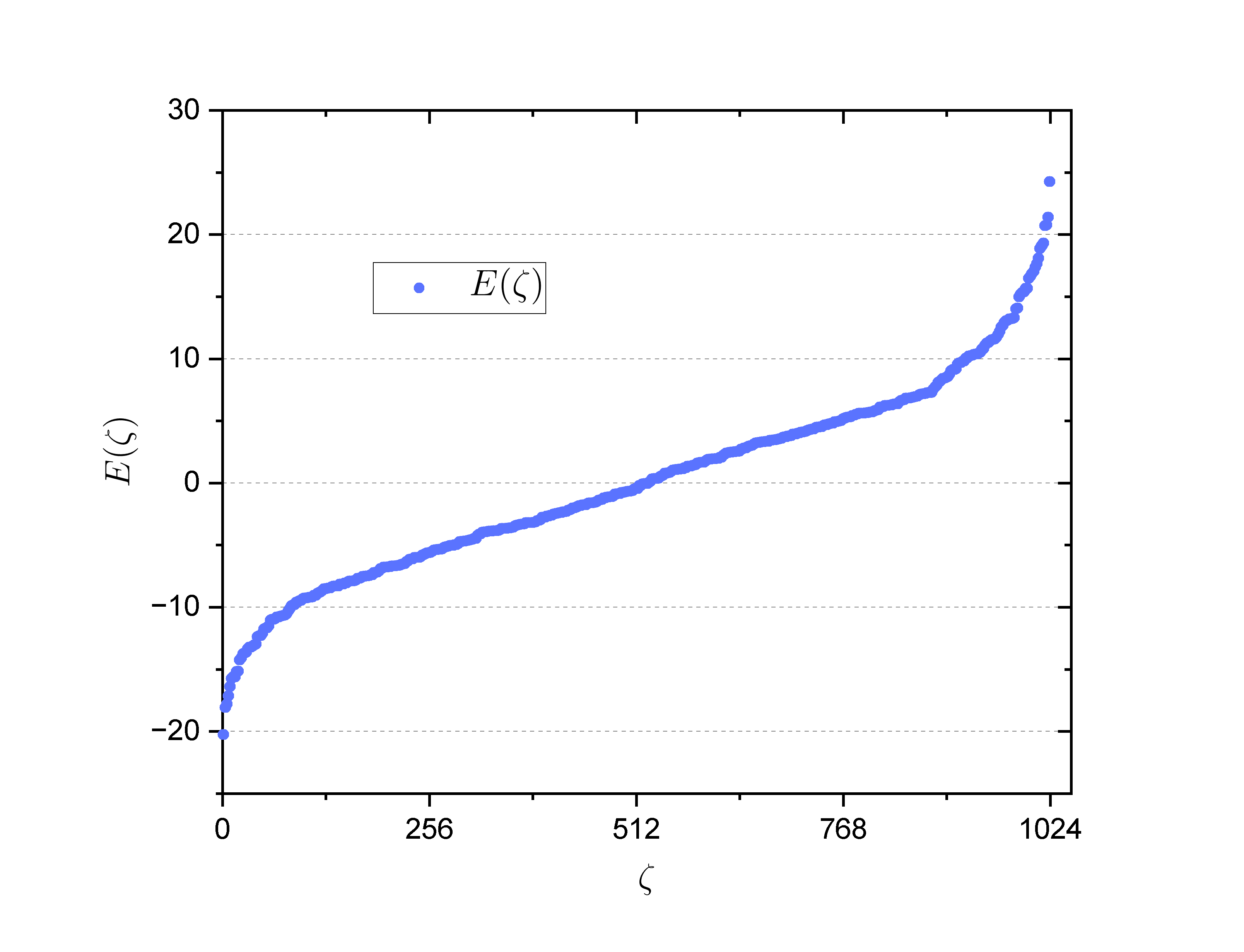}
	}
	\subfigure{(b)}{
		\includegraphics[width=7cm]{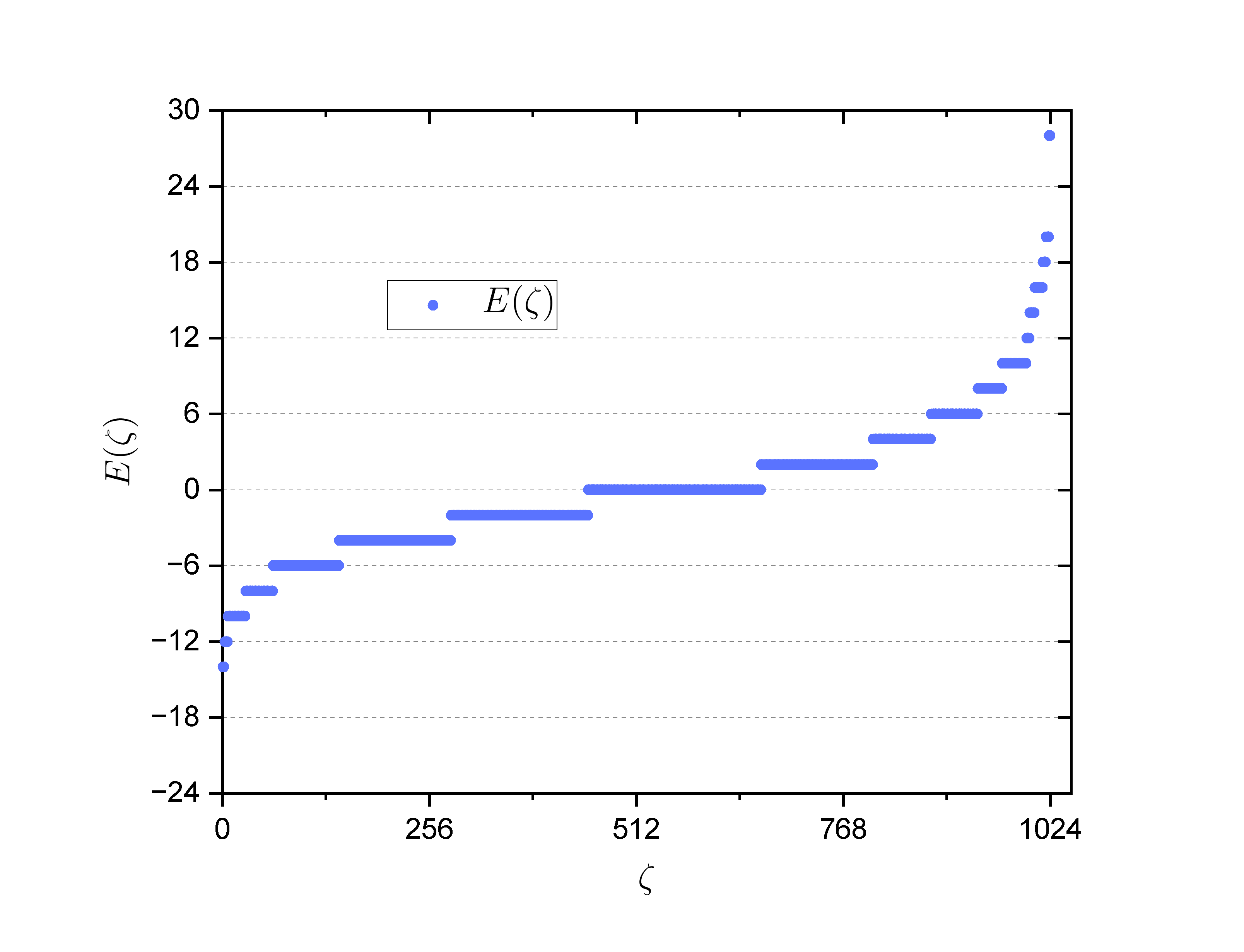}
	}
	\captionsetup{justification=raggedright}
	\caption{\label{fig:costfunction}\sffamily{ We reorganize the energy spectrum by sorting energies in ascending order and assign a energy label $\zeta$ from 0 to 1023 for a clearer illustration of degeneracy. (a) The scatter graph for energy spectrum with $\zeta$ as X-axis about an $n=10$ random instance of Gaussian weighted 2-edge graph. (b) The scatter graph for energy spectrum with $\zeta$ as X-axis about an $n=10$ random instance of MAX-2-XOR, where we choose edge density is 0.6. } }
\end{figure}
	
	In order to illustrate the algorithm performance in application, we start by introducing the two important optimization problems with different degeneracy where we apply our algorithm. The energy spectrum is shown in Figure \ref{fig:costfunction}. The first one is Gaussian weighted 2-edge graph. It is related to Sherrington-Kirkpatrick model\cite{Claes_2021}, and energy spectrum is weak degenerate as Figure \ref{fig:costfunction}(a) shows. The operator corresponding to Gaussian weighted 2-edge graph is
	\begin{equation}\label{eqn-31}
		\mathcal{C} = \sum_{\{i,j\} \in \{1,2,\cdots,n\}} c_{ij} \hat{Z}_{i}\hat{Z}_{j},
	\end{equation}
	where $c_{ij}\sim\mathrm{N}(0,1)$ is without loss of generality, since the results can be promoted to a general Gaussian distribution by the affine property. And Figure \ref{fig:costfunction}(b) is corresponding to MAX-2-XOR, which is one of the most important combinatorial optimization problem. In this paper, we write the operator corresponding to MAX-2-XOR\cite{chou2022limitations}
	\begin{equation}\label{eqn-30}
		\mathcal{C} = \sum_{\{i,j\}\subset\{1,2,\cdots,n\}} d_{i,j} \hat{Z}_{i}\hat{Z}_{j} ,
	\end{equation}
	where each $d_{q_1\cdots q_k}\in\{0,1\}$. We omit the constant term since a constant shift is corresponding to a overall phase which does not influence the objective result. In application, we usually optimize $-\mathcal{C}$ for maximum, which is equivalent to the optimization problem minimizing (\ref{eqn-30}). Especially, we will use the edge density to refer to the frequency that 1 appears in coefficients. The discrete and bounded coefficients result in an much more degenerate spectrum than Gaussian weighted 2-edge graph. We generate the MAX-2-XOR instance in condition that the edge density is set to 0.6.

\begin{figure}[t]
		\centering
		\subfigure{(a)}{
			\includegraphics[width=7cm]{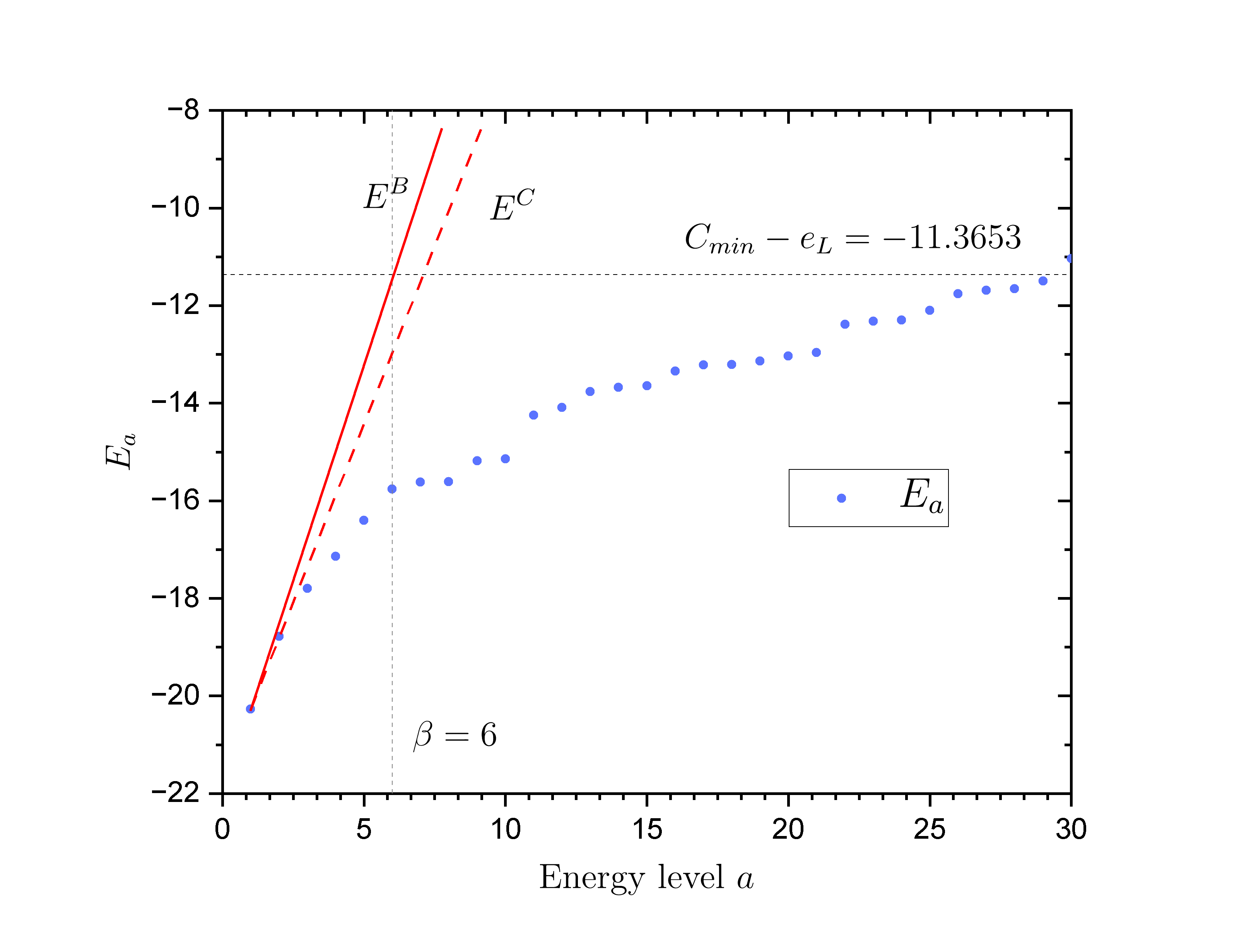}
		}
		\subfigure{(b)}{
			\includegraphics[width=7cm]{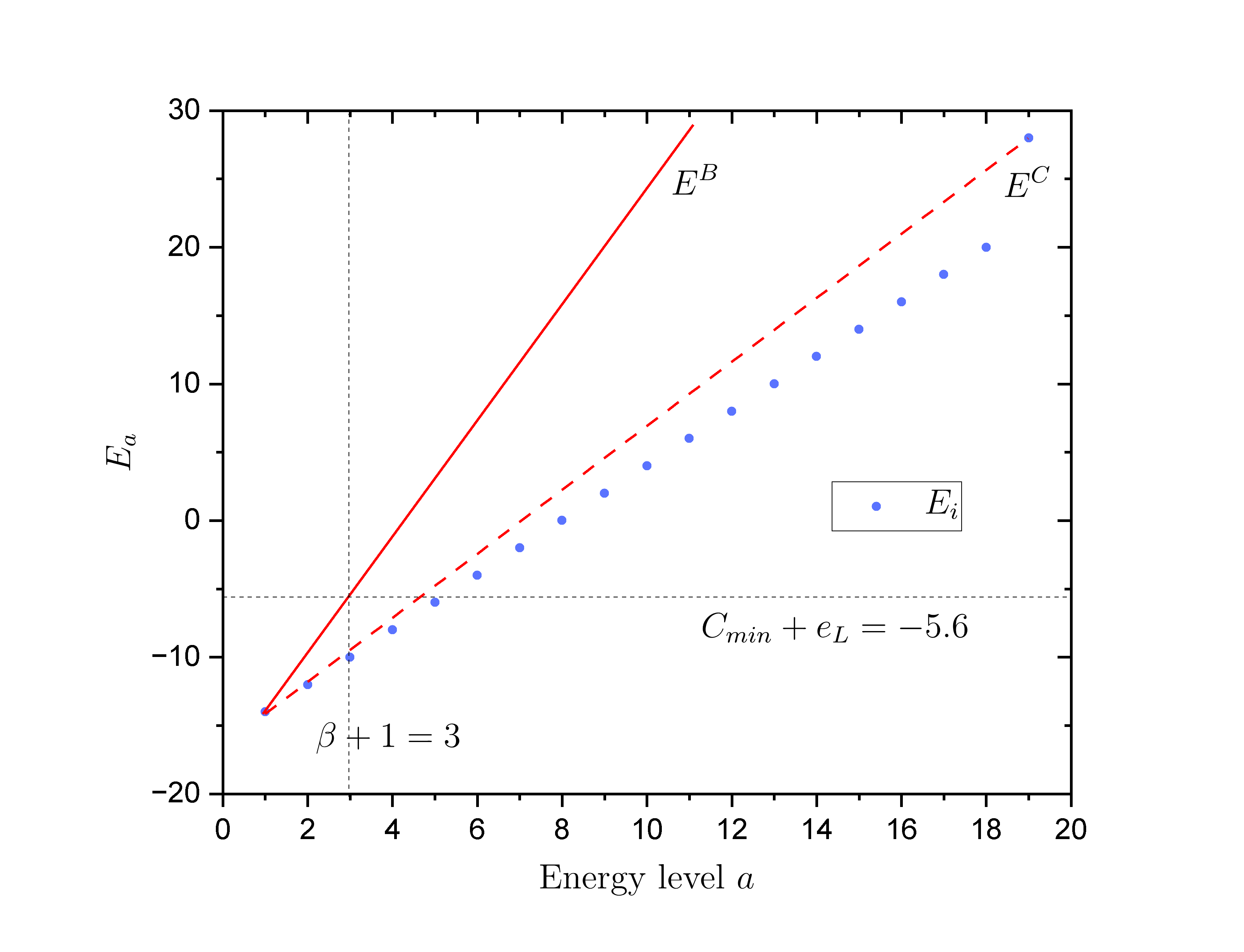}
		}
		\captionsetup{justification=raggedright}
		\caption{\label{fig:upperline_two}\sffamily{ (a) The energy changing graph for upper bound straight line $E^B(a)$, critical straight line $E^C(a)$ and true energy of the Gaussian weight 2-edge graph. Since the total number of energy levels is large, we locally enlarge the energy graph of the first several energy levels.  (b) The energy changing graph for upper bound straight line $E^B(a)$, critical straight line $E^C(a)$ and true energy of the MAX-2-XOR instance.} }
	\end{figure}

	We apply our algorithm to two randomly generated instances belonging to Gaussian weight 2-edge graph and MAX-2-XOR respectively, where the scaled factor $M$ is chosen as $\frac{45}{2\pi}$ and correspondingly $t=7$. Since our $p$-level algorithm use $O(2^p(n+t))$ qubits in the whole procedure, we change $p$ from small to large. Then we obtain the simulation results in form of probability distribution over the computational basis vectors and reorganize these results to calculate the quantities reflecting optimization performance that we introduced in section \ref{c3}. Moreover, we still choose $\epsilon_R= 0.2$, and illustrate the corresponding critical line $E^C(a)$, $E^B(a)$ and true energy changing with energy levels in Figure \ref{fig:upperline_two}(a) and (b) for the two instances respectively.

	 Our results for the Gaussian weight 2-edge graph is shown in Table \ref{tab:gauss}.
	\begin{table}
		\centering
		\begin{threeparttable}
			\centering
			\tabcolsep=0.54cm
			\begin{tabular}{c|c| c |c | c| c| c }
				\hline
				$\ $ & $p$ & $\beta$+1 & $A_{\beta}$ & $\mathcal{P}(U(1,\beta),p)$ & $\bar{\alpha}_R(U(1,\beta))$ & $Pr(\beta+1|U(1,\beta))$   \\ \hline \hline
				$E^B(a)$ & 9  & 6  & 12  &  0.9976 & 0.0226  & 0.0041  \\ \hline
				\multirow{6}{*}{$E_a$} & 9  & \multirow{6}{*}{29}  & \multirow{6}{*}{58}  & 1
				& 0.0164  & $2.040\times10^{-13}$    \\ \cline{2-2} \cline{5-7}
				& 8  &   &   & 1 & 0.0356 &   $2.297\times 10^{-7}$  \\ \cline{2-2} \cline{5-7}
				& 7  &  &   & 1 &  0.0618 & 0.0002    \\ \cline{2-2} \cline{5-7}
				& 6  & & &  0.9760 & 0.0897 & 0.0035   \\ \cline{2-2} \cline{5-7}
				& 5  & & &  0.8452 & 0.1118  & 0.0125 \\ \cline{2-2} \cline{5-7}
				& 4  & & &  0.6066 & 0.1244  & 0.0218   \\ \hline
			\end{tabular}
		\end{threeparttable}
		\captionsetup{justification=raggedright}
		\caption{\label{tab:gauss}\sffamily{ Our calculated value of the cumulative probability over $U(1,\beta)$, the average relative error on $U(1,\beta)$ and the probability proportion that obtain the worst case (the $(\beta+1)$-th energy level) in $U(1,\beta)$ denoted as $Pr(\beta+1|U(1,\beta))$ about the Gaussian weighted 2-edge graph.} }
	\end{table}
	As $p$ increases, the cumulative probability over the energy levels satisfying the worst relative error increases rapidly, which agrees with inequality (\ref{eqn-115}). Users can obtain those near-optimal solutions as expected with considerably higher probability by choosing a larger $p$. Obviously, when $p$ is large enough, those solutions out of $U(1,\beta)$ occupy very low probability that can be ignored in measurement. Thus the algorithm in application shows a considerable reliability. Also, it can be seen that the average relative error on $U(1,\beta)$ obviously decreases. Moreover, we investigate the probability proportion of the worst relative error occurring, and find it decreases in a considerable speed to 0. Combining the cumulative probability converging to 1 quickly, the decreasing reflects that the probability accumulates on those energy levels more optimum. Furtherly, when we consider the optimization problem with low $Pr(\beta+1|U(1,\beta))$ and considerably good $\bar{\alpha}_R(U(1,\beta))$, such as the instance corresponding to Table \ref{tab:gauss}, we can appropriately relax the worst bound $\epsilon_R$, and correspondingly the neighborhood of the ground state $U(1,\beta)$ contains more energy level. In this way, with a lower $p$, we can still obtain the near-optimal solutions according to the new $\epsilon_R$ with expected cumulative probability and acceptable average relative error.
	
	Similarly, the results for the MAX-2-XOR instance is shown in Table \ref{tab:max2xor}. Because of degeneracy, $U(1,\beta)$ consists of more strings than the Gaussian weight 2-edge graph. It can be seen that $\mathcal{P}(U(1,\beta),p)$ and $Pr(\beta+1|U(1,\beta))$ are obviously higher than Table \ref{tab:gauss}. But the trend of performance changing with $p$ increasing is identical in the two problems. We can conclude that the optimization improves steadily and users obtain near-optimal solutions as expected with remarkably higher probability as $p$ increases, no matter what degeneracy the instance is. 
	
	The increasing of $p$ can result in extra resources consumption, especially the spatial complexity, which is proportional to $2^p$. Notably, in both Table \ref{tab:gauss} and Table \ref{tab:max2xor}, we can find that the cumulative probability over energy levels satisfying the worst relative error requirement when $p=5$ is still 0.8452 and 0.9922 respectively. And even when $p=7$, the algorithm ensure the cumulative probability over $U(1,\beta)$ for both of the two instances has been close enough to 1. However, as to $p=9$, the improvement is not worth spending several times more resources just in term of obtaining a solution satisfying the worst relative error. In application, we need to balance the trade-off between the resource consumption and optimization performance.
	\begin{table}
		\centering
		\begin{threeparttable}
			\centering
			\tabcolsep=0.54cm
			\begin{tabular}{c|c| c |c | c| c| c }
				\hline
				$\ $ & $p$ & $\beta$+1 & $A_{\beta}$ & $\mathcal{P}(U(1,\beta),p)$ & $\bar{\alpha}_R(U(1,\beta))$ & $Pr(\beta+1|U(1,\beta))$ \\ \hline \hline
				$E^B(a)$ & 9  & 3  & 28  &  1 & 0.0138  & 0.0024  \\ \hline
				\multirow{6}{*}{$E_a$} & 9  & \multirow{6}{*}{5}  & \multirow{6}{*}{144}  & 1
				& 0.0199  & $1.294\times10^{-14}$  \\ \cline{2-2} \cline{5-7}
				& 8  &   &   & 1 & 0.0395 & $1.138\times 10^{-7}$  \\ \cline{2-2} \cline{5-7}
				& 7  &  &   & 1 &  0.0609 & 0.0003    \\ \cline{2-2} \cline{5-7}
				& 6  & & &  0.9999 & 0.0837 & 0.0183   \\ \cline{2-2} \cline{5-7}
				& 5  & & &  0.9922 & 0.1096  & 0.1287 \\ \cline{2-2} \cline{5-7}
				& 4  & & & 0.9115 & 0.1325  & 0.3068   \\ \hline
			\end{tabular}
		\end{threeparttable}
		\captionsetup{justification=raggedright}
		\caption{\label{tab:max2xor}\sffamily{Our calculated value of the cumulative probability over $U(1,\beta)$, the average relative error on $U(1,\beta)$ and the probability proportion that obtain the worst case in $U(1,\beta)$ about the MAX-2-XOR instance. } }
	\end{table}

	\section{Discussion and conclusion}\label{c5}
	
	In this paper, we propose a pure quantum approximate optimization algorithm. It is different from the known QAOA which relies on the parameters produced by classical method\cite{farhi2014quantum}. Our algorithm is constructed to a $p$-level divide-and-conquer structure based on the CNR operations, as the sketch is shown in Figure \ref{fig:divide_conquer}. The CNR operation can be realized by two steps, comparison and replacement, with the aid of $t$ ancillary qubits. By viewing the Hermitian operator $\mathcal{C}$ as a Hamiltonian, we quantitatively analyze how the CNR operation produces a final state from the initial state in the general form. And the explicit relation is shown in (\ref{eqn-25}). Based on this theoretical result, we derive the problem-independent relation (\ref{eqn-109}), which reveals the core optimization mechanism of our algorithm.
	
	Considering the uncertainty from quantum computation, we introduce the relative error defined in (\ref{eqn-140}) to quantize the quality of approximate optimization. And according to (\ref{eqn-109}), for reasonable $p$, our algorithm finally produces a state in which we can measure and obtain the near-optimal solutions with considerably high probability. Here, by the language introduced in section \ref{c3}, the near-optimal solutions refer to those $z^*$ such that $C(z^*)$ is close to $C_{min}$ within the relative error bound $\epsilon_R$ as required. Particularly, we demonstrate that our algorithm is scalable to optimization problems with different sizes, which ensure the reliability of our algorithm, especially when applied to large-size problems. Moreover, the analysis and discussion on the algorithm performance do not introduce extra requirement on the optimization problem, thus the theoretical conclusion can be applicable in sufficiently general optimization problems. Our $p$-level algorithm requires $O(2^p(n+t))$ qubits including ancillary qubits in $O(p)$ level of simultaneous CNR operations. And the time consumption of the CNR operation is determined by the Quantum Phase Estimation with $t$ ancillary qubits plus an $n$-qubit overwriting operation.
	
	Subsequently, we apply our algorithm to two important optimization problems with different degeneracy, the Gaussian weighted 2-edge graph and MAX-2-XOR, and calculate the algorithm performance and illustrate the results via the quantities that are defined previously. In the actual application, as $p$ increase, the cumulative probability over the near-optimal solutions also converges to 1 very quickly, which agrees with previous theoretical analysis. Moreover, the average relative error decreases steadily. The quality of approximation in actual application improves as expected with the increasing of $p$. However, the spatial resource consumption and required gates are proportional to $2^p$ also results in a rapid expansion when $p$ increases. Thus users face a trade-off between the performance and cost. According to Table \ref{tab:gauss} and Table \ref{tab:max2xor}, we can choose a reasonable $p$ in condition that the cumulative probability does not decrease seriously and the worst relative error is not unacceptable bad to reduce the resources consumption.

	And there are some open questions raised by this work. First, the CNR operation can receive arbitrary superposition state of $2n$-computational basis vectors in principle, not just the uniform superposition. Additionally, the CNR operations within the same level can have different inputs. Thus the possible selections of states stored jointly by the registers have the great degree of freedom. Further works maybe worth discussing the influence of initial states and attempt to give some heuristic choices of inputs for specific optimization problems.
	
	Second, it is worth emphasizing that choosing a suitable $p$ also has a great degree of freedom, but faces a trade-off of efficiency and performance. Though the increasing of $p$ can bring obvious improvements on performance, such as the cumulative probability over near-optimal solutions satisfying the worst bound exponentially converging to 1, we need to consider if the performance improvement is worth consuming much more times resources. Thus it is interesting to study the suitable choosing strategy for $p$ to a specific or even general optimization problems, which is useful in further actual applications.

\begin{acknowledgements}
	This work was supported by National Key R$\&$ D Program of China under Grant No. 2018YFB1601402-2.
\end{acknowledgements}

\bibliography{ref}

\end{document}